\documentclass[twocolumn,showpacs,preprintnumbers,amsmath,amssymb, prl]{revtex4-1}

\usepackage{stmaryrd}
\usepackage{txfonts}
\usepackage{amssymb}
\usepackage{mathrsfs}
\usepackage{graphicx}
\usepackage{dcolumn}
\usepackage{bm}
\usepackage{epsfig}

\begin{document}

\title{Prediction of polaron-like vortices and dissociation depinning transition in magnetic superconductors: The example of ErNi$_2$B$_2$C }

\author{Lev N. Bulaevskii and Shi-Zeng Lin}
\affiliation{Theoretical Division, Los Alamos National Laboratory, Los Alamos, New Mexico 87545}

\begin{abstract}
In borocarbide ErNi$_2$B$_2$C the phase transition to the commensurate spin density wave at 2.3 K
leaves 1/20 part of Ising-like Er spins practically free. Vortices polarize these spins nonuniformly and repolarize them when moving. 
At a low spin relaxation rate and at low bias currents vortices carrying 
magnetic polarization clouds become polaron-like and their velocities are determined by the effective drag coefficient 
which is significantly bigger than the Bardeen-Stephen (BS) one. As current increases, at a critical current $J_c$ vortices release 
polarization clouds and the velocity as well as the voltage in the I-V characteristics jump to values corresponding to the 
BS drag coefficient. The nonuniform components of the magnetic field and magnetization drop as velocity increases 
resulting in weaker polarization and {\it discontinuous} dynamic dissociation depinning transition. 
As current decreases, on the way back, vortices are retrapped by polarization clouds at the current $J_r<J_c$.
 
 \end{abstract}
\date{\today}
\maketitle

The family of quaternary nickel borocarbides (RE)Ni$_2$B$_2$C (RE is rare earth magnetic ion) is an interesting class of crystals which exhibits both superconductivity and magnetic order at low temperatures. \cite{Canfield98,Budko06,Gupta2006}. A number of the crystals in that family develop antiferromagnetic order below the 
N\'{e}el temperature $T_N$ below the superconducting critical temperature $T_c$. It has been recognized time ago that superconductivity
coexists quite peacefully with the antiferromagnetic order as the spatial periodicity of magnetic moments is well below 
the superconducting correlation length. In contrast,  the ferromagnetic order, antagonistic to the Cooper pairing,  leads 
to dramatic changes in both magnetic and superconducting orders in the coexistence phase, for a review see Ref. \cite{Bulaevskii85}. 
That is why interest in the compound ErNi$_2$B$_2$C with $T_c=11$ K and $T_N=6$ K peaked when it was realized 
that below the phase transition from incommensurate spin density wave (SDW) to commensurate SDW at $T^*=2.3$ K the phase with weak ferromagnetic ordering may emerge.~\cite{Cho1995,Canfield1996} From neutron scattering data it was concluded that in ErNi$_2$B$_2$C below $T_N$ the incommensurate SDW develops with effective Ising spins oriented along the $a$-axis and with the wave vector $Q=0.5526b^*$,
where $b^*=2\pi/b$ and $b$ is the lattice period along the $b$-axis.
~\cite{Choi2001,Kawano2002} At $T^*$ the transition to the commensurate phase with $Q=0.55 b^*$ leaves one out of 20 spins free of SDW molecular field. These Er spins with the magnetic moment $\mu=7.8\mu_B$ 
are easily polarizable by the magnetic field along the $a$ direction. The spin magnetization in the magnetic field $H=2000$ G in temperature interval 2 K - 4 K follows $M_{{\rm sp}}/H\approx  \mu M_s/(k_BT)$, where
$M_s\approx 56$ G, see Fig.~4 
in Ref.~\cite{Canfield1996}.
This value, $M_s=\mu n$, corresponds to magnetization when all "free" spins with the concentration $n$ order ferromagnetically,
The same value $M_s$ was obtained by extrapolation of the magnetization at temperature 2 K
in fields $H>1500$ G to $H\rightarrow 0$.  \cite{Gammel2000} 
Note, that the Hall probe measurements below $T^*$ without an applied field found magnetic internal field much lower than $M_s$ and no spontaneous vortex lattice was seen.~\cite{Bluhm2006} High polarizability of spin system in ErNi$_2$B$_2$C is a key point for our following discussion.

As hope to observe remarkable consequences of weak ferromagnetic phase coexisting with superconductivity waned, a few puzzles on
ErNi$_2$B$_2$C behavior at low temperatures remained. First, it was discovered by measuring the hysteresis in the $M-H$ loops 
and transport measurements that new pinning mechanism develops below $T^*$ for which the critical current increases as temperature lowers.~\cite{Gammel2000,James2001} Second, neutron scattering data in applied magnetic field ${\mathbf H}$ 
close to ${\mathbf H}\parallel {\mathbf c}$ have shown that 
vortices deviate randomly from the direction of the magnetic field inside the crystal.~\cite{Yaron1996} The vortex deviations
 increase proportionally to $1/T$ as temperature drops from 4 K to 1.6 K and is nearly independent of the magnetic field in contrast to the usual behavior 
when effect of disorder drops with field.  So far, no explanation of surprising  temperature and field dependence of the critical current and disorder has been offered. 

To explain these data we propose a new mechanism of pinning - formation of polaron-like vortices dressed by the polarization cloud
of magnetic moments. The polaronic mechanism is inherent to all magnetic superconductors but 
it is best pronounced when the magnetic system is highly polarizable, as in  the case of  
ErNi$_2$B$_2$C below 4 K. To clarify this mechanism, we recall 
that the magnetic field is nonuniform within the vortex lattice being strongest near the vortex cores. Consequently, the polarization of the magnetic moments is also nonuniform. When vortices move they should repolarize the magnetic system, otherwise they would lose the energy gained by 
polarization (the Zeeman energy). The 
process of repolarization depends on the dynamics of magnetic system.  In the following we consider the 
relaxation dynamics of free spins in ErNi$_2$B$_2$C. The repolarization process is controlled by the relaxation time $\tau$ which should be compared with the characteristic time 
$a/v$ needed to shift the vortex lattice moving with the velocity $v$ by the vortex lattice period $a=(\Phi_0/B)^{1/2}$ ($\Phi_0$ the flux quantum and $B$ the magnetic induction). For $\tau\gg a/v$ the magnetic moments slow down strongly the vortex motion.
 At some critical velocity and critical current, the vortices are stripped off polarization clouds. The corresponding jump in velocity is strongly pronounced for a large $\tau$'s. As current decreases, the vortices become retrapped again at the current $J_{r}<J_c$. Since the voltage 
$V\propto v$, the I-V characteristics show hysteresis. The physics here is similar to that of a polaron with vortices playing the role of electrons and the magnetic polarization in place of phonons \cite{AppelBook}.
 
\begin{figure}[t]
\psfig{figure=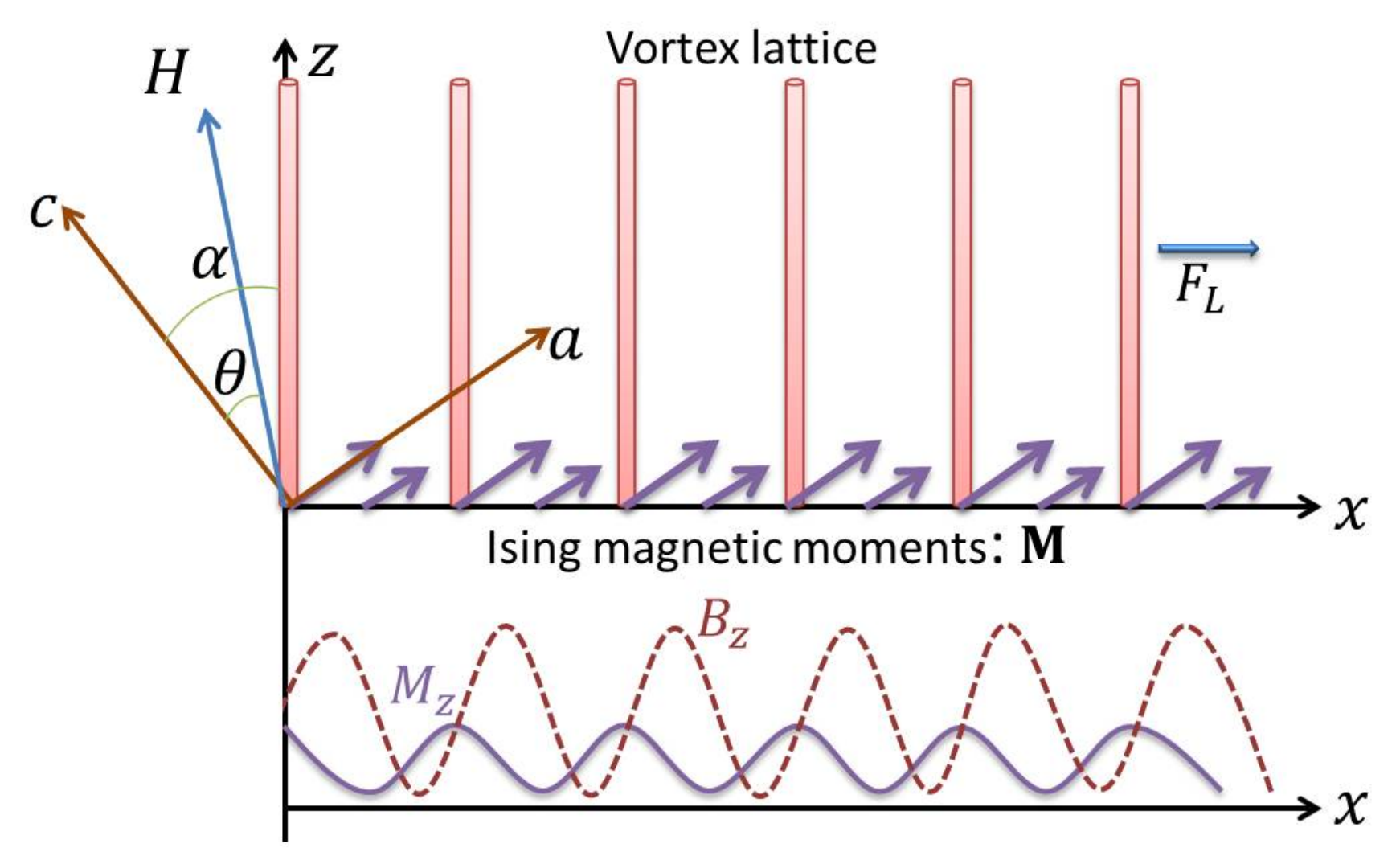,width=\columnwidth}
\caption{\label{f1}(color online) Schematic view of the vortex lattice in the presence of free Ising magnetic moments along the $a$ axis. The vortex lattice is tilted from the applied magnetic fields in the $ac$ plane due to the polarization of the magnetic moments. The vertical columns show the vortex cores. The polarized magnetic moments are nonuniform in space due to the spatial modulation of the vortex lattice magnetic field. Due to the Lorentz force $F_L$ vortices move along the $x$ axis. In moving lattice, there is a phase shift between the magnetic induction (dashed line) associated with the vortex lattice and the magnetization (solid line) caused by the retardation in the response of magnetic moments to the vortex magnetic field.} 
\end{figure}

The ErNi$_2$B$_2$C crystals  have orthorhombic structure below $T_N$ with domains where $a$- and $b$-axes change by $90^\circ$
in neighbouring domains. We consider first clean single-domain crystal and later will discuss the effect of domain walls. 
We consider the vortex lattice induced by applied magnetic field ${\mathbf H}$ tilted by the angle $\theta$ with respect to the crystal $c$ axis. 
We choose the $z$ axis along the direction of vortex lines at rest and $x$ axis in the $ac$-plane, see Fig. \ref{f1}. The vortex line deviates from the applied field $\mathbf{H}$ due to the magnetic moments \cite{Yaron1996}.   In static situation the direction of vortex lines is determined by the effective field ${\mathbf H}+4\pi\overline{{\mathbf M}}$. Here $\overline{{\mathbf M}}$ is the spatial average of the magnetization. We denote by $\alpha$ the angle between vortex lines and the $c$ axis.  
The deviation of vortex lines from the applied magnetic field was also discussed in Ref.~\cite{Ng1997} in the case of  spontaneous ferromagnetic order of the magnetic moments. 

 In the London approximation the magnetic field of the vortex lattice inside the crystal is $({\bf r}=x,y)$ 
\begin{equation}\label{eq2}
B_z({\bf r})=\bar{B}_z\sum_{\mathbf G}\frac{\cos({\mathbf G}\cdot{\mathbf r})}{\lambda^2\mathbf{G}^2+1}, 
\end{equation}
where ${\mathbf G}$ are reciprocal vectors of the square lattice, $\lambda$ is the superconducting penetration length renormalized by the magnetic moments and $\bar{B}$ is the averaged magnetic induction. Here we ignore anisotropy of the penetration length. As revealed by neutron scattering, vortices form square lattice in ErNi$_2$B$_2$C. \cite{Yaron1996}

In the Lagrangian the interaction between vortex lines at ${\mathbf R}_i=(x_i,y_i)$, and the  magnetic moments is determined by the term 
\begin{equation}\label{eq3}
{\cal L}_{\rm{int}}\{{\mathbf R}_i,{\mathbf M}\}=-\int dt\int d{\mathbf r} B_z(\mathbf{R}_i-\mathbf {r},t) M_z({\mathbf r},t),
\end{equation}
where we describe the magnetic moments in the continuous approximation via the magnetization $M_z({\mathbf r},t)$, because distance between free spins, 35 \AA,~\cite{Kawano2002} is much smaller than the London penetration length $\lambda$, about 500 \AA. \cite{Yaron1996}
We ignore the pair breaking effect \cite{Ramakrishnan1981} of the magnetic moments because they suppress Cooper pairing uniformly as distance between free spins is much smaller than the coherence length, and thus moments do not introduce pinning. We also neglect the effect of disorder in crystal lattice. 
The equation of motion for vortex lines is
\begin{equation}\label{eq5}
\eta\frac{\partial {\mathbf R}_i}{\partial t}=-\frac{\partial{\cal L}_{\rm{vv}}\{{\mathbf R}_i,{\mathbf R}_j\}}{\partial \mathbf{R}_i}-\frac{\partial {\cal L}_{\rm{int}}\{{\mathbf R}_i,{\mathbf M}\}}{\partial {\mathbf R}_i}+\mathbf{F}_L,
\end{equation}
where $\eta$ is the Bardeen-Stephen drag coefficient, ${\cal L}_{\rm{vv}}\{{\mathbf R}_i,{\mathbf R}_j\}$ is the vortex-vortex interaction,
next  term describes the force acting on the vortex line ${\bf R}_i$ from the magnetic moments and $F_L=\Phi_0 J/c$ is the Lorentz force due to the bias current $J$. The force due to magnetic moments is the same for all lines and the vortex lattice moves as a whole. The motion of vortex lattice center of mass, $u(t)$, along the $x$-axis is described by the equation 
\begin{equation}\label{eq6}
\eta \frac{\partial u}{\partial t}=\frac{\partial}{\partial u}\left[\int d{\mathbf r}B_z(x+u,y,t)M_z({\mathbf r},t)\right]+F_L,
\end{equation}
Using  linear response approach to relate magnetization with the magnetic field  we obtain
\begin{eqnarray}\label{eq7}
&& \eta \frac{\partial u}{\partial t}=\frac{\partial}{\partial u}\int d{\mathbf r} d{\bf r}'B_z(x+u,y,t)\int_0^tdt'\chi_{zz}({\bf r}-{\bf r}',t-t')\times\nonumber \\
&&B_z({\mathbf r}',t')+F_L,
\end{eqnarray}
Here $\chi_{zz}({\bf r},t)$ is the dynamic susceptibility of the magnetic moments. 
The vortex lattice moves with a constant velocity, $u=vt$, in the steady state $t\gg \tau$. Integrating over coordinates and time we obtain
\begin{equation}\label{eq8}
\eta v=\sum_{{\bf G}}\frac{\chi_{zz}({\bf G}, {\bf v}\cdot{\bf G})}{(\lambda^2{\bf G}^2+1)^{2}}+F_L,
\end{equation}
where $\chi_{zz}({\bf k},\omega)$ is the dynamic magnetic susceptibility in the Fourier representation. We see that the magnetic moments affect strongly vortex motion if a) the resonance condition ${\bf v}\cdot{\bf G}=\Omega({\bf G})$ is fulfilled, where $\Omega({\bf k})$ is the frequency of magnetic excitations with the momentum ${\bf k}$ 
and $\Omega({\bf k})\gg\Gamma({\bf k})$, where $\Gamma({\bf k})$ is the relaxation rate of excitation, 
and b) dynamics of magnetic system is dominated by relaxation, 
$\Omega({\bf k})\lesssim\Gamma({\bf k})$, 
favouring formation of the polaron. 
In the former case, discussed in Ref.~\cite{Shekhter11}, the magnetic moments renormalize the vortex viscosity at high velocities when 
alternating magnetic field of vortices is able to excite magnons. Here we consider the latter case of free moments described by the relaxation dynamics with $\chi_{zz}(\mathbf{k}, \omega)$ given by
\begin{equation}
\chi_{zz}(\mathbf{k}, \omega)=\chi\sin^2\alpha\frac{1}{1-i\omega\tau}, \ \ \ \ \chi=\frac{\mu M_s}{k_BT}
\end{equation}
at temperatures $T$ below 4 K for ErNi$_2$B$_2$C
(effect of oredered spins will be discussed below). 

\begin{figure}[t]
\psfig{figure=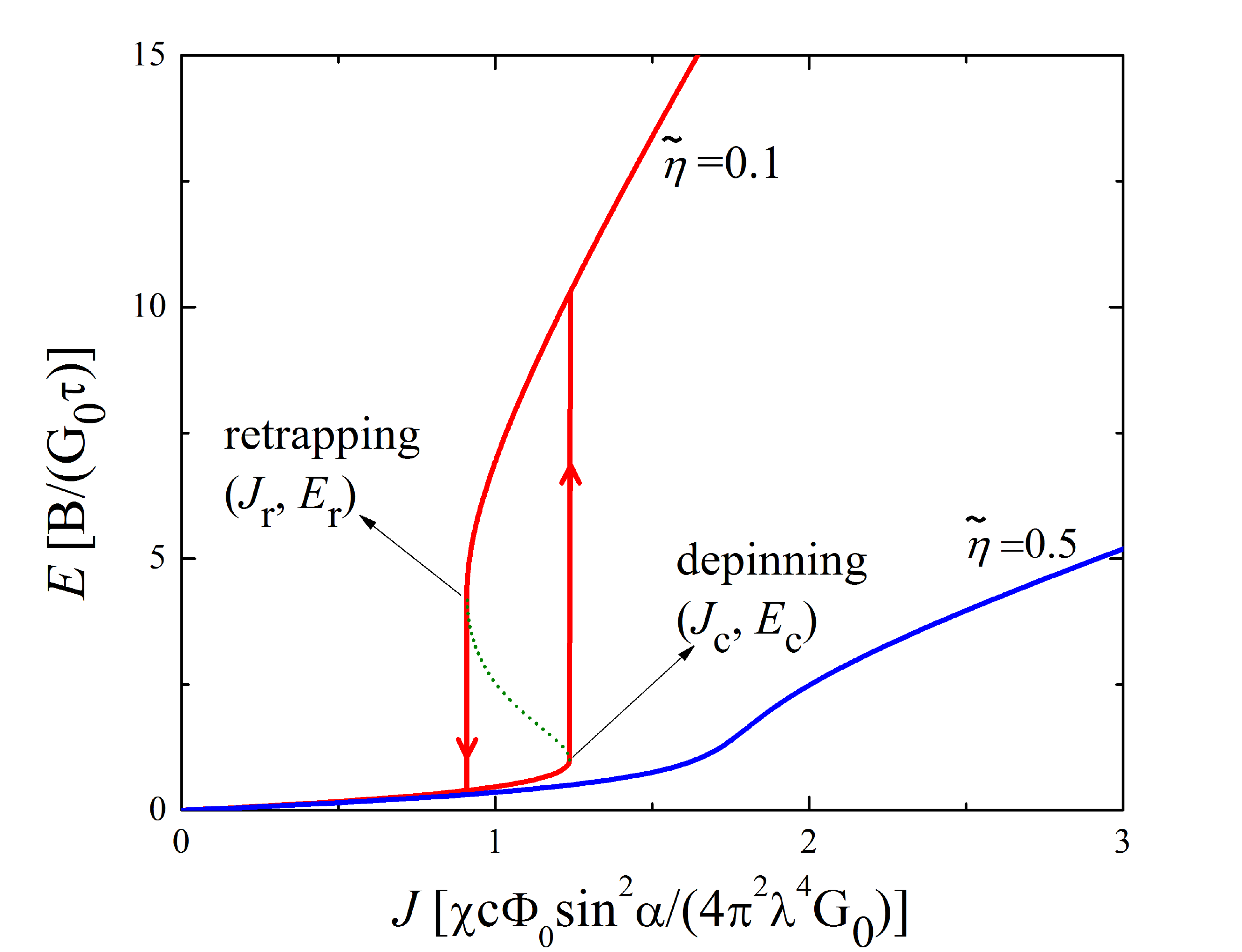,width=\columnwidth}
\caption{\label{f2} Calculated I-V curves for $\tilde{\eta}=0.1$ and $\tilde{\eta}=0.5$. For $\tilde{\eta}=0.1$, the system shows hysteresis in the I-V curve while for $\tilde{\eta}=0.5$, no hysteresis is present. The green dotted line denotes the unstable solution.}
\end{figure}

We introduce dimensionless quantities by expressing $t$ in units of $\tau$ and $u$ in units $G_0^{-1}=a/(2\pi)$. In the summation over $G_y$ in Eq. (\ref{eq8}), we account only for the dominant term with $G_y=0$.
For $G_0\lambda\gg 1$ we find  equation for the dimensionless vortex velocity
\begin{equation}\label{eq10}
\tilde{\eta}v=F(v)+\tilde{F}_L, \ \ \ \         F(v)=-v   [\pi^2/3+v^2-\pi v\coth(\pi/v)], 
\end{equation}
where we introduced the dimensionless parameters  
\begin{equation}\label{eq9}
\tilde{\eta}=\frac{4\pi^2\eta\lambda^4}{{\chi}\Phi_0^2\tau\sin^2\alpha}, \ \ \ \ \tilde{F}_L=\frac{4\pi^2G_0\lambda^4 J}{{\chi} c\Phi_0\sin^2\alpha}.
\end{equation}
The asymptotic behaviour of $F(v)$ is $F(v)\approx -v$ at $v\ll 1$ and $F(v)\approx -\pi^4/(45v)$ at $v\gg 1$. The electric field due to the motion of the vortex lattice is given by $E=B v/c$, and we obtain the I-V curves from Eq.~(\ref{eq10}), as depicted in Fig.~\ref{f2}. As $v$ increases, the $v-J$ curve changes from a weak current dependence $v=\tilde{F}_L/(1+\tilde{\eta})$ to a stronger and usual Bardeen-Stephen behaviour $v=\tilde{F}_L/\tilde{\eta}$. In Fig.~\ref{f2} at $\tilde{\eta}=0.1$,  the I-V curve is hysteretic. Upon ramping up the bias current, the system jumps to the usual Bardeen-Stephen (BS) Ohmic curve at a current $J_c$, where electric field increases discontinuously by the factor $1/\tilde{\eta}$ at $\tilde{\eta}\ll1$. 
The jump, identifying experimentally as depinning transition, is caused by the dissociation of the vortex-magnon polaron.
 It is very similar to 
 the dissociation of usual electron-phonon polaron in high electric fields as described theoretically \cite{Banyai1993} and confirmed experimentally in metal oxides \cite{Hed1970}. Upon decreasing the current the vortices are retrapped by the polarization clouds at a threshold current $J_r$ and the vortex lattice moves with a significantly enhanced viscosity at lower currents.

\begin{figure}[b]
\psfig{figure=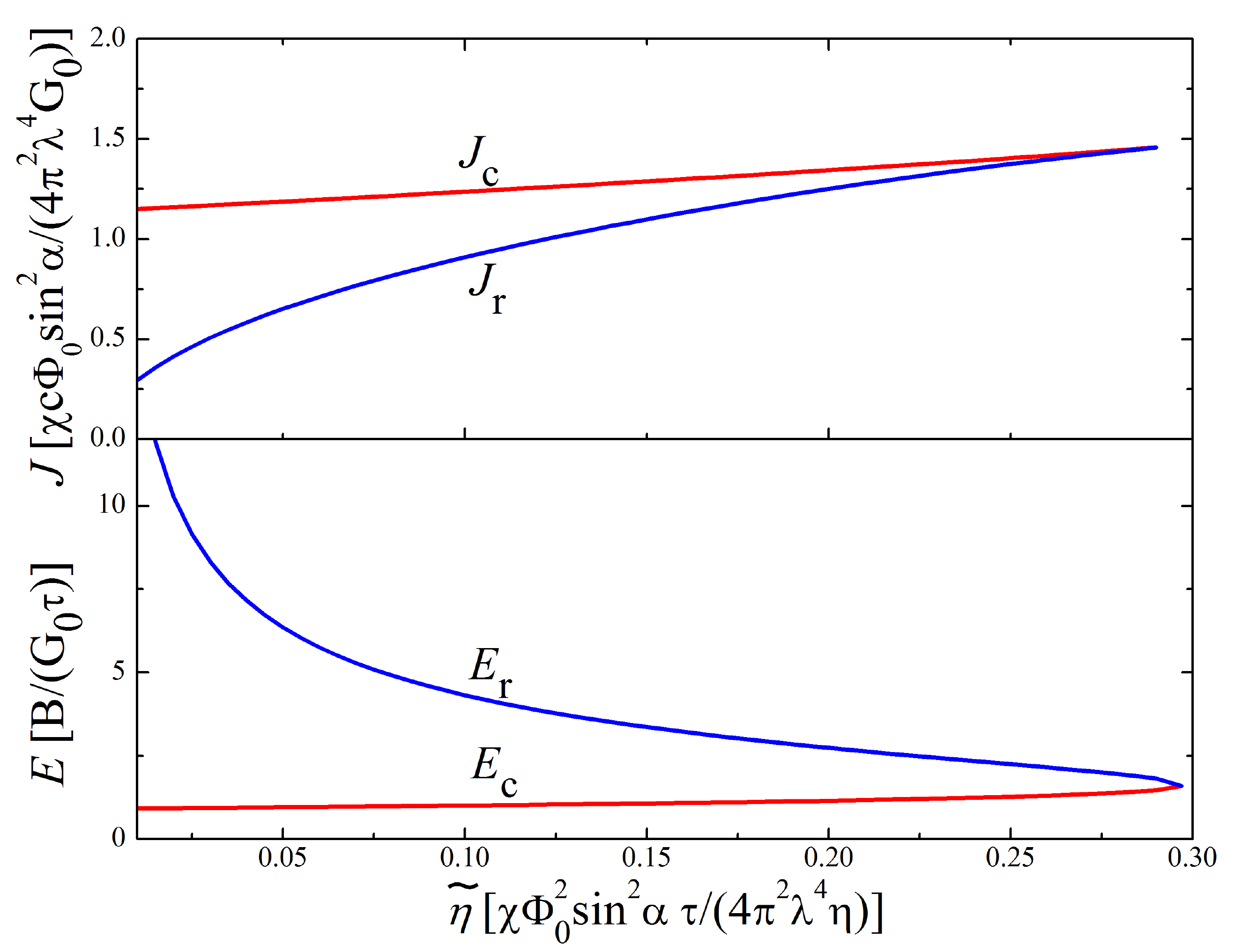,width=\columnwidth}
\caption{\label{f3} Dependence of the critical current $J_c$ and  retrapping current $J_r$, and corresponding electric fields $E_c$ and $E_r$ on $\tilde{\eta}$.}
\end{figure}

The critical current $J_c$ and retrapping current $J_r$ follow from the equation for the velocity $\tilde{\eta}-d F(v)/d v=0$. The maximum of $d F(v)/d v$ is $0.297$, thus hysteresis exists for $\tilde{\eta}<0.297$.  The calculated $J_c$, $J_r$ and corresponding electric fields are shown in Fig. \ref{f3}.  At small $\tilde{\eta}$ the critical current is
\begin{equation}
J_c\approx 0.03\frac{\chi c\Phi_0\sin^2\alpha}{ G_0 \lambda^4}.  
\label{cc}\end{equation}
$J_c$ decreases with temperature as $J_c\sim 1/T$ and decreases with the magnetic field as $J_c\sim 1/\sqrt{B}$. 


We note, that above $T^*$, in the 
incommensurate SDW, some spins experience quite weak SDW molecular field. Thus, they are polarized by vortices and exhibit polaronic effect and pinning. This explains increase of pinning in ErNi$_2$B$_2$C as $T$ decreases below $T_N$, see Ref.~\cite{Gammel2000}, as well as pinning in the holmium borocarbide below $T_N$. ~\cite{Dewhurst1999}

Let us consider the origin of the jumps at $J_c$ and $J_r$. The dependence of the magnetization on the  
velocity of moving vortices is
\begin{equation}\label{eq10a}
{M_z({\bf r},v, t)}={\chi\bar{B}\sin^2\alpha}\sum_{{\bf G}}\frac{\cos[ \mathbf{G}\cdot\mathbf{r}-\beta(v)]}{(\lambda^2{\bf G}^2+1)[1+ (G_x v \tau)^2]^{3/2}}
\end{equation}
with $\tan(\beta)=G_x v\tau$. Nonuniform component of the magnetization and thus the polarization effect decrease with velocity. On the other hand, the retardation between the magnetic field and the magnetization, as described by the phase shift $\beta(v)$, increases with the velocity. This positive feedback and the increase of retardation with velocity ensure discontinuous transitions at $J_c$ and $J_r$.

Strong pinning due to the polaron mechanism requires small parameter $\tilde{\eta}$. It is expressed via 
$\tau$ as $\tilde{\eta}\approx 10^{-11} \rm{s}/(\tau\sin^2\alpha)$, where we have used the BS drag coefficient $\eta_{BS}=\Phi _0^2/(2\pi  \xi ^2c^2\rho_n)$ with the coherence length $\xi\approx 13\ \rm{nm}$ \cite{Yaron1996} and the normal resistivity  $\rho_n=5~\mu\Omega\cdot$cm at $T_c$. \cite{Cava1994} The relaxation time $\tau$ in ErNi$_2$B$_2$C is long because the dynamics of majority of spins is strongly suppressed by the formation of the SDW molecular field as was found by the M\"{o}ssbauer measurements.\cite{Bonville1996} The relaxation time drops very fast below 10 K and reaches the value      
$\tau\approx 5\times 10^{-10}$ s at $T=5$ K, but data at lower temperatures were not reported. Thus the only information we have so far is $\tilde{\eta}<0.02/\sin^2\alpha$. 

The critical current for ErNi$_2$B$_2$C reported in Ref.~\cite{Gammel2000}  for $B=0.1\ \rm{T}$, $T=2\ \rm{K}$ is about 250 A/cm$^2$. 
 Eq.~(\ref{cc}) gives such current value at $\alpha=2.5^{\circ}$. 
 In experiment the applied magnetic field 
was close to the $c$ axis, but the precise angle $\theta$ was not reported.~\cite{Gammel2000}. The estimate of the order 1$^{\circ}$ is reasonable, but quantitative comparison is not convincing as we do not know $\tau$ and thus $\tilde{\eta}$ below 2.3 K. Hence, the real check of polaronic mechanism should be by measuring the I-V characteristics.
We predict hysteretic behavior in ErNi$_2$B$_2$C, strong dependence of voltage and of the critical current on the angle $\theta$, 
at least for $\theta\gg 0.15$. Note, that the critical current reaches values as high as 10$^6$ A/cm$^2$ at high angles at $T=1$ K and $B=0.1$ T. 

The effect of ordered spins on the vortex motion is
similar to that described in Ref. \onlinecite{Shekhter11} for an antiferromagnet. When Cherenkov condition 
${\bf v}\cdot{\bf G}\geq \Omega({\bf G})$ is met, excitation of magnons results in enhanced viscosity $\eta$. This occurs at high velocities, due to a gap in the magnon spectrum, and thus at high currents $J>J_c$, leading to a voltage drop in comparison with the BS result.

Let us discuss now the effect of disorder on the vortex lines direction observed in magnetic fields tilted with respect to the $c$-axis. \cite{Yaron1996}
Due to domain structure the Ising spins are polarized only in half domains where vortex lines follow direction of the effective field ${\mathbf H}+4\pi {\mathbf M}$, while in others they are along ${\mathbf H}$. The random change of angle of vortex directions with respect to average angle 
is $2\pi M/B=2\pi\chi\sin\alpha$. It increases as $1/T$, when $T$ drops, in agreement with the results of Ref.~\cite{Yaron1996} and data for $M_{{\rm sp}}/H$ mentioned above.
When vortices cross a domain wall between domains with different magnetization, they need to repolarize magnetic moments at currents below the critical one. This slows additionally vortex motion, but such an effect is smaller than that accounted for previously because domain size is much bigger than the distance between the magnetic moments. Domain walls also cause pinning of vortices as seen in Bitter decoration patterns.~\cite{Veschunov2007} Importantly, for a nonzero $\theta$ small part of vortices experience domain wall pinning and this part drops with $\theta$. In contrast, polarization effect increases with $\theta$ and this helps to separate domain wall pinning from the polarization one.

In conclusions, vortices in magnetic superconductors polarize magnetic moments and become dressed and polaron-like. At low currents and long spin relaxation time the nonuniform polarization induced by vortices slows their motion at currents for which pinning by crystal lattice disorder becomes ineffective. As current increases above  the critical one,  vortices release nonuniform part of the polarization and the velocity as well as the voltage in the I-V characteristics jump to much higher values. At decreasing current vortices are retrapped by polarized magnetic moments at the retrapping current which is smaller than the critical one. The results of such polaronic mechanism are in qualitative agreement with the experimental data \cite{Yaron1996,Gammel2000} but measurements of the I-V characteristics are needed to establish the quantitative agreement and confirm validity of such a model for Er borocarbide. The polaronic mechanism should be at play also in Gd and Tb borocarbides superconductors in the commensurate SDW phase. 
It may be present in Tm borocarbide above $T_N$ and in cuprate superconductors (RE)Ba$_2$Cu$_3$O$_7$, where magnetic RE ions positioned between superconducting layers 
interact weakly with superconducting electrons and order at 
very low N\'{e}el temperatures of the order 1 K. \cite{Allenspach1995}

\acknowledgments
The authors would like to thank P. Canfield, C. Batista, V. Kogan, D. Smith, A. Saxena and B. Maiorov for helpful discussions.
This publication was made possible by funding from the Los Alamos Laboratory Directed Research and Development Program, project number 20110138ER.


\begin{thebibliography}{23}%
\makeatletter
\providecommand \@ifxundefined [1]{%
 \@ifx{#1\undefined}
}%
\providecommand \@ifnum [1]{%
 \ifnum #1\expandafter \@firstoftwo
 \else \expandafter \@secondoftwo
 \fi
}%
\providecommand \@ifx [1]{%
 \ifx #1\expandafter \@firstoftwo
 \else \expandafter \@secondoftwo
 \fi
}%
\providecommand \natexlab [1]{#1}%
\providecommand \enquote  [1]{``#1''}%
\providecommand \bibnamefont  [1]{#1}%
\providecommand \bibfnamefont [1]{#1}%
\providecommand \citenamefont [1]{#1}%
\providecommand \href@noop [0]{\@secondoftwo}%
\providecommand \href [0]{\begingroup \@sanitize@url \@href}%
\providecommand \@href[1]{\@@startlink{#1}\@@href}%
\providecommand \@@href[1]{\endgroup#1\@@endlink}%
\providecommand \@sanitize@url [0]{\catcode `\\12\catcode `\$12\catcode
  `\&12\catcode `\#12\catcode `\^12\catcode `\_12\catcode `\%12\relax}%
\providecommand \@@startlink[1]{}%
\providecommand \@@endlink[0]{}%
\providecommand \url  [0]{\begingroup\@sanitize@url \@url }%
\providecommand \@url [1]{\endgroup\@href {#1}{\urlprefix }}%
\providecommand \urlprefix  [0]{URL }%
\providecommand \Eprint [0]{\href }%
\providecommand \doibase [0]{http://dx.doi.org/}%
\providecommand \selectlanguage [0]{\@gobble}%
\providecommand \bibinfo  [0]{\@secondoftwo}%
\providecommand \bibfield  [0]{\@secondoftwo}%
\providecommand \translation [1]{[#1]}%
\providecommand \BibitemOpen [0]{}%
\providecommand \bibitemStop [0]{}%
\providecommand \bibitemNoStop [0]{.\EOS\space}%
\providecommand \EOS [0]{\spacefactor3000\relax}%
\providecommand \BibitemShut  [1]{\csname bibitem#1\endcsname}%
\let\auto@bib@innerbib\@empty
\bibitem [{\citenamefont {Canfield}\ \emph {et~al.}(1998)\citenamefont
  {Canfield}, \citenamefont {Gammel},\ and\ \citenamefont
  {Bishop}}]{Canfield98}%
  \BibitemOpen
  \bibfield  {author} {\bibinfo {author} {\bibfnamefont {P.~C.}\ \bibnamefont
  {Canfield}}, \bibinfo {author} {\bibfnamefont {P.~L.}\ \bibnamefont
  {Gammel}}, \ and\ \bibinfo {author} {\bibfnamefont {D.~J.}\ \bibnamefont
  {Bishop}},\ }\href@noop {} {\bibfield  {journal} {\bibinfo  {journal} {Phys.
  Today}\ }\textbf {\bibinfo {volume} {51}},\ \bibinfo {pages} {40} (\bibinfo
  {year} {1998})}\BibitemShut {NoStop}%
\bibitem [{\citenamefont {Bud'ko}\ and\ \citenamefont
  {Canfield}(2006)}]{Budko06}%
  \BibitemOpen
  \bibfield  {author} {\bibinfo {author} {\bibfnamefont {S.~L.}\ \bibnamefont
  {Bud'ko}}\ and\ \bibinfo {author} {\bibfnamefont {P.~C.}\ \bibnamefont
  {Canfield}},\ }\href@noop {} {\bibfield  {journal} {\bibinfo  {journal} {C.
  R. Physique}\ }\textbf {\bibinfo {volume} {7}},\ \bibinfo {pages} {56}
  (\bibinfo {year} {2006})}\BibitemShut {NoStop}%
\bibitem [{\citenamefont {Gupta}(2006)}]{Gupta2006}%
  \BibitemOpen
  \bibfield  {author} {\bibinfo {author} {\bibfnamefont {L.~C.}\ \bibnamefont
  {Gupta}},\ }\href@noop {} {\bibfield  {journal} {\bibinfo  {journal} {Adv.
  Phys.}\ }\textbf {\bibinfo {volume} {55}},\ \bibinfo {pages} {691} (\bibinfo
  {year} {2006})}\BibitemShut {NoStop}%
\bibitem [{\citenamefont {Bulaevskii}\ \emph {et~al.}(1985)\citenamefont
  {Bulaevskii}, \citenamefont {Buzdin}, \citenamefont {Kulic},\ and\
  \citenamefont {Panjukov}}]{Bulaevskii85}%
  \BibitemOpen
  \bibfield  {author} {\bibinfo {author} {\bibfnamefont {L.~N.}\ \bibnamefont
  {Bulaevskii}}, \bibinfo {author} {\bibfnamefont {A.~I.}\ \bibnamefont
  {Buzdin}}, \bibinfo {author} {\bibfnamefont {M.~L.}\ \bibnamefont {Kulic}}, \
  and\ \bibinfo {author} {\bibfnamefont {S.~V.}\ \bibnamefont {Panjukov}},\
  }\href@noop {} {\bibfield  {journal} {\bibinfo  {journal} {Adv. Phys.}\
  }\textbf {\bibinfo {volume} {34}},\ \bibinfo {pages} {175} (\bibinfo {year}
  {1985})}\BibitemShut {NoStop}%
\bibitem [{\citenamefont {Cho}\ \emph {et~al.}(1995)\citenamefont {Cho},
  \citenamefont {Canfield}, \citenamefont {Miller}, \citenamefont {Johnston},
  \citenamefont {Beyermann},\ and\ \citenamefont {Yatskar}}]{Cho1995}%
  \BibitemOpen
  \bibfield  {author} {\bibinfo {author} {\bibfnamefont {B.~K.}\ \bibnamefont
  {Cho}}, \bibinfo {author} {\bibfnamefont {P.~C.}\ \bibnamefont {Canfield}},
  \bibinfo {author} {\bibfnamefont {L.~L.}\ \bibnamefont {Miller}}, \bibinfo
  {author} {\bibfnamefont {D.~C.}\ \bibnamefont {Johnston}}, \bibinfo {author}
  {\bibfnamefont {W.~P.}\ \bibnamefont {Beyermann}}, \ and\ \bibinfo {author}
  {\bibfnamefont {A.}~\bibnamefont {Yatskar}},\ }\href@noop {} {\bibfield
  {journal} {\bibinfo  {journal} {Phys. Rev. B}\ }\textbf {\bibinfo {volume}
  {52}},\ \bibinfo {pages} {3684} (\bibinfo {year} {1995})}\BibitemShut
  {NoStop}%
\bibitem [{\citenamefont {Canfield}\ \emph {et~al.}(1996)\citenamefont
  {Canfield}, \citenamefont {Bud'ko},\ and\ \citenamefont
  {Cho}}]{Canfield1996}%
  \BibitemOpen
  \bibfield  {author} {\bibinfo {author} {\bibfnamefont {P.}~\bibnamefont
  {Canfield}}, \bibinfo {author} {\bibfnamefont {S.}~\bibnamefont {Bud'ko}}, \
  and\ \bibinfo {author} {\bibfnamefont {B.}~\bibnamefont {Cho}},\ }\href@noop
  {} {\bibfield  {journal} {\bibinfo  {journal} {Physica C}\ }\textbf {\bibinfo
  {volume} {262}},\ \bibinfo {pages} {249} (\bibinfo {year}
  {1996})}\BibitemShut {NoStop}%
\bibitem [{\citenamefont {Choi}\ \emph {et~al.}(2001)\citenamefont {Choi},
  \citenamefont {Lynn}, \citenamefont {Lopez}, \citenamefont {Gammel},
  \citenamefont {Canfield},\ and\ \citenamefont {Bud'ko}}]{Choi2001}%
  \BibitemOpen
  \bibfield  {author} {\bibinfo {author} {\bibfnamefont {S.}~\bibnamefont
  {Choi}}, \bibinfo {author} {\bibfnamefont {J.~W.}\ \bibnamefont {Lynn}},
  \bibinfo {author} {\bibfnamefont {D.}~\bibnamefont {Lopez}}, \bibinfo
  {author} {\bibfnamefont {P.~L.}\ \bibnamefont {Gammel}}, \bibinfo {author}
  {\bibfnamefont {P.~C.}\ \bibnamefont {Canfield}}, \ and\ \bibinfo {author}
  {\bibfnamefont {S.~L.}\ \bibnamefont {Bud'ko}},\ }\href@noop {} {\bibfield
  {journal} {\bibinfo  {journal} {Phys. Rev. Lett.}\ }\textbf {\bibinfo
  {volume} {87}},\ \bibinfo {pages} {107001} (\bibinfo {year}
  {2001})}\BibitemShut {NoStop}%
\bibitem [{\citenamefont {{Kawano-Furukawa}}\ \emph {et~al.}(2002)\citenamefont
  {{Kawano-Furukawa}}, \citenamefont {Takeshita}, \citenamefont {Ochiai},
  \citenamefont {Nagata}, \citenamefont {Yoshizawa}, \citenamefont {Furukawa},
  \citenamefont {Takeya},\ and\ \citenamefont {Kadowaki}}]{Kawano2002}%
  \BibitemOpen
  \bibfield  {author} {\bibinfo {author} {\bibfnamefont {H.}~\bibnamefont
  {{Kawano-Furukawa}}}, \bibinfo {author} {\bibfnamefont {H.}~\bibnamefont
  {Takeshita}}, \bibinfo {author} {\bibfnamefont {M.}~\bibnamefont {Ochiai}},
  \bibinfo {author} {\bibfnamefont {T.}~\bibnamefont {Nagata}}, \bibinfo
  {author} {\bibfnamefont {H.}~\bibnamefont {Yoshizawa}}, \bibinfo {author}
  {\bibfnamefont {N.}~\bibnamefont {Furukawa}}, \bibinfo {author}
  {\bibfnamefont {H.}~\bibnamefont {Takeya}}, \ and\ \bibinfo {author}
  {\bibfnamefont {K.}~\bibnamefont {Kadowaki}},\ }\href@noop {} {\bibfield
  {journal} {\bibinfo  {journal} {Phys. Rev. B}\ }\textbf {\bibinfo {volume}
  {65}},\ \bibinfo {pages} {180508} (\bibinfo {year} {2002})}\BibitemShut
  {NoStop}%
\bibitem [{\citenamefont {Gammel}\ \emph {et~al.}(2000)\citenamefont {Gammel},
  \citenamefont {Barber}, \citenamefont {Lopez}, \citenamefont {Ramirez},
  \citenamefont {Bishop}, \citenamefont {Bud'ko},\ and\ \citenamefont
  {Canfield}}]{Gammel2000}%
  \BibitemOpen
  \bibfield  {author} {\bibinfo {author} {\bibfnamefont {P.~L.}\ \bibnamefont
  {Gammel}}, \bibinfo {author} {\bibfnamefont {B.}~\bibnamefont {Barber}},
  \bibinfo {author} {\bibfnamefont {D.}~\bibnamefont {Lopez}}, \bibinfo
  {author} {\bibfnamefont {A.~P.}\ \bibnamefont {Ramirez}}, \bibinfo {author}
  {\bibfnamefont {D.~J.}\ \bibnamefont {Bishop}}, \bibinfo {author}
  {\bibfnamefont {S.~L.}\ \bibnamefont {Bud'ko}}, \ and\ \bibinfo {author}
  {\bibfnamefont {P.~C.}\ \bibnamefont {Canfield}},\ }\href@noop {} {\bibfield
  {journal} {\bibinfo  {journal} {Phys. Rev. Lett.}\ }\textbf {\bibinfo
  {volume} {84}},\ \bibinfo {pages} {2497} (\bibinfo {year}
  {2000})}\BibitemShut {NoStop}%
\bibitem [{\citenamefont {Bluhm}\ \emph {et~al.}(2006)\citenamefont {Bluhm},
  \citenamefont {Sebastian}, \citenamefont {Guikema}, \citenamefont {Fisher},\
  and\ \citenamefont {Moler}}]{Bluhm2006}%
  \BibitemOpen
  \bibfield  {author} {\bibinfo {author} {\bibfnamefont {H.}~\bibnamefont
  {Bluhm}}, \bibinfo {author} {\bibfnamefont {S.~E.}\ \bibnamefont
  {Sebastian}}, \bibinfo {author} {\bibfnamefont {J.~W.}\ \bibnamefont
  {Guikema}}, \bibinfo {author} {\bibfnamefont {I.~R.}\ \bibnamefont {Fisher}},
  \ and\ \bibinfo {author} {\bibfnamefont {K.~A.}\ \bibnamefont {Moler}},\
  }\href@noop {} {\bibfield  {journal} {\bibinfo  {journal} {Phys. Rev. B}\
  }\textbf {\bibinfo {volume} {73}},\ \bibinfo {pages} {014514} (\bibinfo
  {year} {2006})}\BibitemShut {NoStop}%
\bibitem [{\citenamefont {James}\ \emph {et~al.}(2001)\citenamefont {James},
  \citenamefont {Dewhurst}, \citenamefont {Field}, \citenamefont {Paul},
  \citenamefont {Paltiel}, \citenamefont {Shtrikman}, \citenamefont {Zeldov},\
  and\ \citenamefont {Campbell}}]{James2001}%
  \BibitemOpen
  \bibfield  {author} {\bibinfo {author} {\bibfnamefont {S.~S.}\ \bibnamefont
  {James}}, \bibinfo {author} {\bibfnamefont {C.~D.}\ \bibnamefont {Dewhurst}},
  \bibinfo {author} {\bibfnamefont {S.~B.}\ \bibnamefont {Field}}, \bibinfo
  {author} {\bibfnamefont {D.~M.}\ \bibnamefont {Paul}}, \bibinfo {author}
  {\bibfnamefont {Y.}~\bibnamefont {Paltiel}}, \bibinfo {author} {\bibfnamefont
  {H.}~\bibnamefont {Shtrikman}}, \bibinfo {author} {\bibfnamefont
  {E.}~\bibnamefont {Zeldov}}, \ and\ \bibinfo {author} {\bibfnamefont {A.~M.}\
  \bibnamefont {Campbell}},\ }\href@noop {} {\bibfield  {journal} {\bibinfo
  {journal} {Phys. Rev. B}\ }\textbf {\bibinfo {volume} {64}},\ \bibinfo
  {pages} {092512} (\bibinfo {year} {2001})}\BibitemShut {NoStop}%
\bibitem [{\citenamefont {Yaron}\ \emph {et~al.}(1996)\citenamefont {Yaron},
  \citenamefont {Gammel}, \citenamefont {Ramirez}, \citenamefont {Huse},
  \citenamefont {Bishop}, \citenamefont {Goldman}, \citenamefont {Stassis},
  \citenamefont {Canfield}, \citenamefont {Mortensen},\ and\ \citenamefont
  {Eskildsen}}]{Yaron1996}%
  \BibitemOpen
  \bibfield  {author} {\bibinfo {author} {\bibfnamefont {U.}~\bibnamefont
  {Yaron}}, \bibinfo {author} {\bibfnamefont {P.~L.}\ \bibnamefont {Gammel}},
  \bibinfo {author} {\bibfnamefont {A.~P.}\ \bibnamefont {Ramirez}}, \bibinfo
  {author} {\bibfnamefont {D.~A.}\ \bibnamefont {Huse}}, \bibinfo {author}
  {\bibfnamefont {D.~J.}\ \bibnamefont {Bishop}}, \bibinfo {author}
  {\bibfnamefont {A.~I.}\ \bibnamefont {Goldman}}, \bibinfo {author}
  {\bibfnamefont {C.}~\bibnamefont {Stassis}}, \bibinfo {author} {\bibfnamefont
  {P.~C.}\ \bibnamefont {Canfield}}, \bibinfo {author} {\bibfnamefont
  {K.}~\bibnamefont {Mortensen}}, \ and\ \bibinfo {author} {\bibfnamefont
  {M.~R.}\ \bibnamefont {Eskildsen}},\ }\href@noop {} {\bibfield  {journal}
  {\bibinfo  {journal} {Nature}\ }\textbf {\bibinfo {volume} {382}},\ \bibinfo
  {pages} {236} (\bibinfo {year} {1996})}\BibitemShut {NoStop}%
\bibitem [{\citenamefont {Appel}(1968)}]{AppelBook}%
  \BibitemOpen
  \bibfield  {author} {\bibinfo {author} {\bibfnamefont {J.}~\bibnamefont
  {Appel}},\ }\href@noop {} {\emph {\bibinfo {title} {Polarons, Solid State
  Physics, Editor: F. Seitz and D. Turnbull, vol. 21, p. 193}}}\ (\bibinfo
  {publisher} {Academic Press},\ \bibinfo {address} {New York},\ \bibinfo
  {year} {1968})\BibitemShut {NoStop}%
\bibitem [{\citenamefont {Ng}\ and\ \citenamefont {Varma}(1997)}]{Ng1997}%
  \BibitemOpen
  \bibfield  {author} {\bibinfo {author} {\bibfnamefont {T.~K.}\ \bibnamefont
  {Ng}}\ and\ \bibinfo {author} {\bibfnamefont {C.~M.}\ \bibnamefont {Varma}},\
  }\href@noop {} {\bibfield  {journal} {\bibinfo  {journal} {Phys. Rev. Lett.}\
  }\textbf {\bibinfo {volume} {78}},\ \bibinfo {pages} {3745} (\bibinfo {year}
  {1997})}\BibitemShut {NoStop}%
\bibitem [{\citenamefont {Ramakrishnan}\ and\ \citenamefont
  {Varma}(1981)}]{Ramakrishnan1981}%
  \BibitemOpen
  \bibfield  {author} {\bibinfo {author} {\bibfnamefont {T.~V.}\ \bibnamefont
  {Ramakrishnan}}\ and\ \bibinfo {author} {\bibfnamefont {C.~M.}\ \bibnamefont
  {Varma}},\ }\href@noop {} {\bibfield  {journal} {\bibinfo  {journal} {Phys.
  Rev. B}\ }\textbf {\bibinfo {volume} {24}},\ \bibinfo {pages} {137} (\bibinfo
  {year} {1981})}\BibitemShut {NoStop}%
\bibitem [{\citenamefont {Shekhter}\ \emph {et~al.}(2011)\citenamefont
  {Shekhter}, \citenamefont {Bulaevskii},\ and\ \citenamefont
  {Batista}}]{Shekhter11}%
  \BibitemOpen
  \bibfield  {author} {\bibinfo {author} {\bibfnamefont {A.}~\bibnamefont
  {Shekhter}}, \bibinfo {author} {\bibfnamefont {L.~N.}\ \bibnamefont
  {Bulaevskii}}, \ and\ \bibinfo {author} {\bibfnamefont {C.~D.}\ \bibnamefont
  {Batista}},\ }\href@noop {} {\bibfield  {journal} {\bibinfo  {journal} {Phys.
  Rev. Lett.}\ }\textbf {\bibinfo {volume} {106}},\ \bibinfo {pages} {037001}
  (\bibinfo {year} {2011})}\BibitemShut {NoStop}%
\bibitem [{\citenamefont {B\'{a}nyai}(1993)}]{Banyai1993}%
  \BibitemOpen
  \bibfield  {author} {\bibinfo {author} {\bibfnamefont {L.}~\bibnamefont
  {B\'{a}nyai}},\ }\href@noop {} {\bibfield  {journal} {\bibinfo  {journal}
  {Phys. Rev. Lett.}\ }\textbf {\bibinfo {volume} {70}},\ \bibinfo {pages}
  {1674} (\bibinfo {year} {1993})}\BibitemShut {NoStop}%
\bibitem [{\citenamefont {Hed}\ and\ \citenamefont {Freud}(1970)}]{Hed1970}%
  \BibitemOpen
  \bibfield  {author} {\bibinfo {author} {\bibfnamefont {A.}~\bibnamefont
  {Hed}}\ and\ \bibinfo {author} {\bibfnamefont {P.}~\bibnamefont {Freud}},\
  }\href@noop {} {\bibfield  {journal} {\bibinfo  {journal} {J. Non-Cryst.
  Solids}\ }\textbf {\bibinfo {volume} {2}},\ \bibinfo {pages} {484} (\bibinfo
  {year} {1970})}\BibitemShut {NoStop}%
\bibitem [{\citenamefont {Dewhurst}\ \emph {et~al.}(1999)\citenamefont
  {Dewhurst}, \citenamefont {Doyle}, \citenamefont {Zeldov},\ and\
  \citenamefont {McK.~Paul}}]{Dewhurst1999}%
  \BibitemOpen
  \bibfield  {author} {\bibinfo {author} {\bibfnamefont {C.~D.}\ \bibnamefont
  {Dewhurst}}, \bibinfo {author} {\bibfnamefont {R.~A.}\ \bibnamefont {Doyle}},
  \bibinfo {author} {\bibfnamefont {E.}~\bibnamefont {Zeldov}}, \ and\ \bibinfo
  {author} {\bibfnamefont {D.}~\bibnamefont {McK.~Paul}},\ }\href@noop {}
  {\bibfield  {journal} {\bibinfo  {journal} {Phys. Rev. Lett.}\ }\textbf
  {\bibinfo {volume} {82}},\ \bibinfo {pages} {827} (\bibinfo {year}
  {1999})}\BibitemShut {NoStop}%
\bibitem [{\citenamefont {Cava}\ \emph {et~al.}(1994)\citenamefont {Cava},
  \citenamefont {Takagi}, \citenamefont {Zandbergen}, \citenamefont
  {Krajewski}, \citenamefont {Peck}, \citenamefont {Siegrist}, \citenamefont
  {Batlogg}, \citenamefont {Dover}, \citenamefont {Felder}, \citenamefont
  {Mizuhashi}, \citenamefont {Lee}, \citenamefont {Eisaki},\ and\ \citenamefont
  {Uchida}}]{Cava1994}%
  \BibitemOpen
  \bibfield  {author} {\bibinfo {author} {\bibfnamefont {R.~J.}\ \bibnamefont
  {Cava}}, \bibinfo {author} {\bibfnamefont {H.}~\bibnamefont {Takagi}},
  \bibinfo {author} {\bibfnamefont {H.~W.}\ \bibnamefont {Zandbergen}},
  \bibinfo {author} {\bibfnamefont {J.~J.}\ \bibnamefont {Krajewski}}, \bibinfo
  {author} {\bibfnamefont {W.~F.}\ \bibnamefont {Peck}}, \bibinfo {author}
  {\bibfnamefont {T.}~\bibnamefont {Siegrist}}, \bibinfo {author}
  {\bibfnamefont {B.}~\bibnamefont {Batlogg}}, \bibinfo {author} {\bibfnamefont
  {R.~B.~v.}\ \bibnamefont {Dover}}, \bibinfo {author} {\bibfnamefont {R.~J.}\
  \bibnamefont {Felder}}, \bibinfo {author} {\bibfnamefont {K.}~\bibnamefont
  {Mizuhashi}}, \bibinfo {author} {\bibfnamefont {J.~O.}\ \bibnamefont {Lee}},
  \bibinfo {author} {\bibfnamefont {H.}~\bibnamefont {Eisaki}}, \ and\ \bibinfo
  {author} {\bibfnamefont {S.}~\bibnamefont {Uchida}},\ }\href@noop {}
  {\bibfield  {journal} {\bibinfo  {journal} {Nature}\ }\textbf {\bibinfo
  {volume} {367}},\ \bibinfo {pages} {252} (\bibinfo {year}
  {1994})}\BibitemShut {NoStop}%
\bibitem [{\citenamefont {Bonville}\ \emph {et~al.}(1996)\citenamefont
  {Bonville}, \citenamefont {Hodges}, \citenamefont {Vaast}, \citenamefont
  {Alleno}, \citenamefont {Godart}, \citenamefont {Gupta}, \citenamefont
  {Hossain}, \citenamefont {Nagarajan}, \citenamefont {Hilscher},\ and\
  \citenamefont {Michor}}]{Bonville1996}%
  \BibitemOpen
  \bibfield  {author} {\bibinfo {author} {\bibfnamefont {P.}~\bibnamefont
  {Bonville}}, \bibinfo {author} {\bibfnamefont {J.~A.}\ \bibnamefont
  {Hodges}}, \bibinfo {author} {\bibfnamefont {C.}~\bibnamefont {Vaast}},
  \bibinfo {author} {\bibfnamefont {E.}~\bibnamefont {Alleno}}, \bibinfo
  {author} {\bibfnamefont {C.}~\bibnamefont {Godart}}, \bibinfo {author}
  {\bibfnamefont {L.~C.}\ \bibnamefont {Gupta}}, \bibinfo {author}
  {\bibfnamefont {Z.}~\bibnamefont {Hossain}}, \bibinfo {author} {\bibfnamefont
  {R.}~\bibnamefont {Nagarajan}}, \bibinfo {author} {\bibfnamefont
  {G.}~\bibnamefont {Hilscher}}, \ and\ \bibinfo {author} {\bibfnamefont
  {H.}~\bibnamefont {Michor}},\ }\href@noop {} {\bibfield  {journal} {\bibinfo
  {journal} {Z. Phys. B}\ }\textbf {\bibinfo {volume} {101}},\ \bibinfo {pages}
  {511} (\bibinfo {year} {1996})}\BibitemShut {NoStop}%
\bibitem [{\citenamefont {Veschunov}\ \emph {et~al.}(2007)\citenamefont
  {Veschunov}, \citenamefont {Vinnikov}, \citenamefont {Bud’ko},\ and\
  \citenamefont {Canfield}}]{Veschunov2007}%
  \BibitemOpen
  \bibfield  {author} {\bibinfo {author} {\bibfnamefont {I.~S.}\ \bibnamefont
  {Veschunov}}, \bibinfo {author} {\bibfnamefont {L.~Y.}\ \bibnamefont
  {Vinnikov}}, \bibinfo {author} {\bibfnamefont {S.~L.}\ \bibnamefont
  {Bud’ko}}, \ and\ \bibinfo {author} {\bibfnamefont {P.~C.}\ \bibnamefont
  {Canfield}},\ }\href@noop {} {\bibfield  {journal} {\bibinfo  {journal}
  {Phys. Rev. B}\ }\textbf {\bibinfo {volume} {76}},\ \bibinfo {pages} {174506}
  (\bibinfo {year} {2007})}\BibitemShut {NoStop}%
\bibitem [{\citenamefont {Allenspach}\ \emph {et~al.}(1995)\citenamefont
  {Allenspach}, \citenamefont {Lee}, \citenamefont {Gajewski}, \citenamefont
  {Barbeta}, \citenamefont {Maple}, \citenamefont {Nieva}, \citenamefont {Yoo},
  \citenamefont {Kramer}, \citenamefont {{McCallum}},\ and\ \citenamefont
  {{Ben-Dor}}}]{Allenspach1995}%
  \BibitemOpen
  \bibfield  {author} {\bibinfo {author} {\bibfnamefont {P.}~\bibnamefont
  {Allenspach}}, \bibinfo {author} {\bibfnamefont {B.~W.}\ \bibnamefont {Lee}},
  \bibinfo {author} {\bibfnamefont {D.~A.}\ \bibnamefont {Gajewski}}, \bibinfo
  {author} {\bibfnamefont {V.~B.}\ \bibnamefont {Barbeta}}, \bibinfo {author}
  {\bibfnamefont {M.~B.}\ \bibnamefont {Maple}}, \bibinfo {author}
  {\bibfnamefont {G.}~\bibnamefont {Nieva}}, \bibinfo {author} {\bibfnamefont
  {S.~I.}\ \bibnamefont {Yoo}}, \bibinfo {author} {\bibfnamefont {M.~J.}\
  \bibnamefont {Kramer}}, \bibinfo {author} {\bibfnamefont {R.~W.}\
  \bibnamefont {{McCallum}}}, \ and\ \bibinfo {author} {\bibfnamefont
  {L.}~\bibnamefont {{Ben-Dor}}},\ }\href@noop {} {\bibfield  {journal}
  {\bibinfo  {journal} {Z. Phys. B}\ }\textbf {\bibinfo {volume} {96}},\
  \bibinfo {pages} {455} (\bibinfo {year} {1995})}\BibitemShut {NoStop}%
\end{thebibliography}
%

\end{document}